\documentclass[doublecol]{epl2}

\title{Subordination model of anomalous diffusion leading to the two-power-law relaxation responses}
\shorttitle{Subordination model of anomalous diffusion \dots}

\author{Aleksander Stanislavsky\inst{1}, Karina Weron\inst{2} \and Justyna Trzmiel\inst{2}}
\shortauthor{A. Stanislavsky \etal}

\institute{
  \inst{1} Institute of Radio Astronomy - 4 Chervonopraporna St., 61002 Kharkov, Ukraine\\
  \inst{2} Institute of Physics,  Wroc{\l}aw University of Technology - Wyb.
Wyspia$\acute{n}$kiego 27, 50-370 Wroc{\l}aw, Poland }

\abstract{ We derive a general pattern of the nonexponential,
two-power-law relaxation from the compound subordination theory of
random processes applied to anomalous diffusion. The subordination
approach is based on a coupling between the very large jumps in
physical and operational times. It allows one to govern a scaling
for small and large times independently. Here we obtain explicitly
the relaxation function, the kinetic equation and the
susceptibility expression applicable to the range of
experimentally  observed power-law exponents which cannot be
interpreted by means of the commonly known Havriliak-Negami
fitting function. We present a novel two-power relaxation law for
this range in a convenient frequency-domain form and show its
relationship to the Havriliak-Negami one.}

\begin{document}

\maketitle

\section{Introduction}
Many studies have been reported on the phenomenon of
nonexponential, power-law relaxation which is typically observed
in complex systems such as dielectrics, ferroelectrics, polymers
and so on (see \cite{J1,J2} and references therein). The main
feature of such systems is a strong (in general, random)
interaction between their components in the passage to a state of
equilibrium. Therefore, theoretical construction of an
``averaged'' object representing the entire relaxing system is not
a simple problem; it requires application of advanced
probabilistic tools. One of them uses a randomization in the
parameters of distributions that describes the relaxation rates in
disordered systems. With regard to the dielectric relaxation, each
individual dipole in a dielectric system relaxes exponentially,
but their relaxation rates are different and obey a probability
distribution (continuous function) \cite{JurWer}. This method is
successive for getting many empirical response laws and their
classification, but it sometimes becomes enough complicated to
interpret their interrelations and to derive kinetic/diffusion
equations. An alternative approach, applied to analysis of
diffusion processes underlying the macroscopic dynamics, is based
on subordination of random processes (see, for example,
\cite{mk2004} and references therein). The anomalous diffusion
process $Y[U(t)]$ is obtained by the time clock randomization of a
random (parent) process $Y(t)$ by means of another random process
$U(t)$ called the directing process. The latter process is also
often referred to as the randomized or operational time
\cite{Feller}. As it has been shown in \cite{magwer06,Stan07}, the
subordination method is also useful for analysis of the relaxation
processes. However, up to now the approach had some difficulties
in interpretation of the Havriliak-Negami (HN) law, one of the
most general description of relaxation data. Only recently the
effective picture of the HN relaxation, based on subordination,
has been found in \cite{wjmwt2010}. It requires introduction of
compound subordinators to the study of the anomalous diffusion.
The compound subordination procedure relates one subordinator to
another in such a way that the temporal evolution of the parent
random process has a different scaling on short and long times.

\begin{figure}
\onefigure[width=8.6 cm]{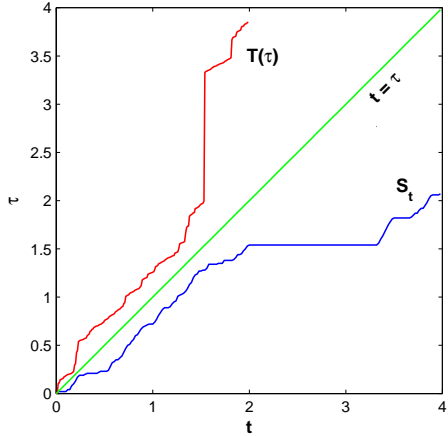} \caption{(Color online)
Trajectory of $T(\tau)$ and $S_t$ with
$\alpha=0.75$.}\label{fig:us}
\end{figure}

\section{Coupling between the $\alpha$-stable process as the real time
and its inverse as an operational  time} The inverse
$\alpha$-stable process is actually the left-inverse process of
the $\alpha$-stable process (see Fig.~\ref{fig:us}). This means
that the $\alpha$-stable process $T(\tau)$ and its inverse
$S_t=\inf\{\tau\geq 0\,|\,T(\tau) >t\}$ fulfill the relation
$S_{T(\tau)}=\tau$, while $T(S_t)>t$ holds. The random variable
$S_t$ corresponds to the first passage time of the strictly
increasing process $T(\tau)$ above $t$. To find the probability
density distribution (pdf) of the compound process $T(S_t)$, we
start with a sum of independent and identically distributed
heavy-tailed random variables $T_i$, namely $U_n=\sum_{i=0}^nT_i$
with $T_0=0$. The process $N_t=\max\{n\in {\bf N}\,|\,U_n\leq t\}$
is inverse to $U_n$. From this definition it follows directly the
inequality
\begin{equation}
U_{N_t}\leq t < U_{N_t+1}\quad {\rm for}\quad t\geq
0\,.\label{eq1}
\end{equation}
According to the Feller's book \cite{Feller}, the renewal theory
tells us how the pdfs of $U_{N_t}$ and $U_{N_t+1}$ behave in the
long time limit.  The random variable $U_{N_t}/t$ tends (in
distribution) to $Y$ with the pdf equal
\begin{equation}
p^Y(x)=\frac{\sin(\pi\alpha)}{\pi}\,x^{\alpha-1}(1-x)^{-\alpha}\,,\quad
0<x<1\,, \label{eq2}
\end{equation}
and $U_{N_t+1}/t$ tends to $Z$ which pdf reads
\begin{equation}
p^Z(x)=\frac{\sin(\pi\alpha)}{\pi}\,x^{-1}(x-1)^{-\alpha}\,,\quad
x>1.\label{eq3}
\end{equation}
The functions $p^Y(x)$ and $p^Z(x)$ correspond to special cases of
the well-known beta density \cite{Feller}. We note that the pdf of
$p^Y(x)$ concentrates near 0 and 1, whereas $p^Z(x)$ concentrates
near 1, where both tend to infinity. The moments of the random
value $Y$ are finite and can be calculated directly from the pdf
(\ref{eq2}). However, even the first moment of $Z$ is infinite.

A passage from the discrete process $T_i$ to the continuous limit
$T(\tau)$ allows one to rewrite the inequality (\ref{eq1}) into a
form
\begin{equation}
T^-(S_t)\leq t < T(S_t)\quad {\rm for}\quad t\geq 0\,,\label{eq4}
\end{equation}
underestimating or overestimating the real time $t$. In this case
we obtain an analogous to (\ref{eq2}) and (\ref{eq3}) result. In
the long limit the pdfs of $T^-(S_t)$ and  $T(S_t)$ read
\begin{eqnarray}
p^-(t,y)&=&
\frac{\sin\pi\alpha}{\pi}\,y^{\alpha-1}(t-y)^{-\alpha}\,,\quad
0<y<t\,,\label{eq5}\\
p^+(t,z)&=&\frac{\sin\pi\alpha}{\pi}\,y^{-1}\,t^\alpha
(y-t)^{-\alpha}\,,\quad y>t\,.\label{eq6}
\end{eqnarray}
Their moments are computed directly from the moments of $Y$ and
$Z$ due to the relations $T^-(S_t)\stackrel{\,d}{=}tY$ and
$T(S_t)\stackrel{\,d}{=}tZ$, where $\stackrel{\,d}{=}$ denotes the
equality in distribution. Thus, the process $T^-(S_t)$ has finite
moments of any order, while $T(S_t)$ gives us even no finite the
first moment. It should be noticed that the processes $T^-(S_t)$
and $T(S_t)$ evolve to infinity with growing $t$. They show a
coupling between the very large jumps in physical and operational
times.

\section{Universality of relaxation in two-state systems} The
simplest interpretation of relaxation processes uses the
concept of a system of independent exponentially relaxing objects
(for example, dipoles) with different (independent) relaxation
rates \cite{Bord78}. The systems, following this law (called
Debye's), can easily be described as a two-state system. Let $N$
be the common number of dipoles in a dielectric system. If
$N_\uparrow$ is the number of dipoles in the state $\uparrow$,
$N_\downarrow$ is the number of dipoles in the state $\downarrow$
so that $N=N_\uparrow+N_\downarrow$. Assume that for $t=0$ the
system is stated in order so that the states $\uparrow$ dominate,
namely
$$\frac{N_\uparrow(t=0)}{N}=n_\uparrow(0)=1,\quad
\frac{N_\downarrow(t=0)}{N}=n_\downarrow(0)=0\,,$$
where $n_\uparrow$ is the ratio of dipoles in the state $\uparrow$
and $n_\downarrow$ in the state $\downarrow$. Denote the
transition rate by $w$ defined from microscopic properties of the
system (for instance, according to given Hamiltonian of
interaction and the Fermi's golden rule). In this case the kinetic
equation is of the form
\begin{equation}
\cases{\dot n_\uparrow(t)-w\,\{n_\downarrow(t)-
n_\uparrow(t)\}=0,&\cr \dot n_\downarrow(t)-w\,\{n_\uparrow(t)-
n_\downarrow(t)\}=0,&\cr}
\label{eq7}
\end{equation}
where the dotted symbol  means the first-order time derivative.
The relaxation function for the two-state system reads then
$\phi_{\rm D}(t)=1-2n_\downarrow(t)=2n_\uparrow(t)-1=\exp(-2wt)$.
It is easy see that the steady state of the system corresponds to
equilibrium with $n_\uparrow(\infty)=n_\downarrow(\infty)=1/2$.
However, if the dipoles interact with their environment, and the
interaction is complex (random), their contribution in relaxation
already will not result in any exponential decay.

Assume that the interaction of dipoles with environment is taken
into account with a help of subordination in time. Take the
process $S_t$ as a subordinator. It accounts for the amount of
time when a dipole does not participate in motion. The ratio of
dipoles in the state $\uparrow$ and  another in the state
$\downarrow$ is subordinated by the process $S_t$. The equation
describing the two-state system takes the form similar to
Eq.(\ref{eq7}), but the derivatives of first order become
fractional of order $0<\alpha<1$ determined by the index of the
process $S_t$ inverse to the $\alpha$-stable process $T(\tau)$.
This leads to the Cole-Cole (CC) relaxation (see
\cite{magwer06,Stan07} for details). The relaxation function for
the two-state system satisfies now the following equation
$$\frac{\partial^\alpha}{\partial t^\alpha}\phi_{\rm CC}(t)-
\frac{t^{-\alpha}}{\Gamma(1-\alpha)} =-\omega_p^\alpha \phi_{\rm
CC}(t)\,,$$
with the initial condition $\phi_{\rm CC}(0)=1$. Here we use the
Riemann-Liouville definition of fractional derivative \cite{Goren}, namely
$$\frac{\partial^\alpha}{\partial t^\alpha}
x(t)=\frac{d^n}{dt^n}\left[\frac{1}{\Gamma(n-\alpha)}\int^t_0\frac{x(\tau)}
{(t-\tau)^{\alpha+1-n}}\,d\tau\right]\,,$$
where $n-1<\alpha<n$. The constant $\omega_p^\alpha$ characterizes
the transition rate from microscopic properties of the system. The
relaxation function reads $\phi_{\rm
CC}(t)=E_\alpha(-\omega_p^\alpha t^\alpha)$, where
$E_\alpha(z)=\sum_{k=0}^\infty z^k/\Gamma(1+k\alpha)$ is the
one-parameter Mittag-Leffler function \cite{Erd}.

For the experimental study the frequency-domain representation of
the latter function
\begin{equation}
\varphi^*(\omega)\propto\int^\infty_0e^{-i\omega
t}\,\left(-\frac{d\phi(t)}{dt}\right)\,dt\label{eq8}
\end{equation}
is of interest. It is well known that the complex dielectric
susceptibility
$\chi(\omega)=\chi'(\omega)-i\chi''(\omega)\propto\varphi^*(\omega)$
of most dipolar substances demonstrates a peak in the loss
component $\chi''(\omega)$ at a characteristic frequency
$\omega_p$. The CC  susceptibility is
\begin{equation}
\chi_{\rm
CC}(\omega)=\frac{1}{1+(i\omega/\omega_p)^\alpha}\,,\quad
0<\alpha\leq 1\,. \label{eq9}
\end{equation}
With reference to the theory of subordination the CC law shows
that the dipoles tend to equilibrium via motion alternating with
stops so that the temporal intervals between them are random.
However, there are other well-know laws of relaxation, in
particular, the Cole-Davidson (CD) and Havriliak-Negami (HN) ones.
The description of the laws requires a modification in the theory
of subordination.

It should be noticed that the physical mechanism underlying the nonexponential
relaxation can be described as a diffusive limit of continuous time random walks.
The resulting relaxation patterns are connected not only with stochastic features of
the jumps and the inter-jump times themselves, but also with a stochastic dependence
between them. In the framework of Linear Response Theory the temporal decay
of a given mode $k$, representing excitation undergoing diffusion in the system
under consideration, is given by the inverse Fourier transform of the diffusion
front \cite{km2000}.

\section{Subordination by the process $T(S_t)$} The subordinator
$T(S_t)$ results in stretching of the real time $t$. It will
underline scaling properties in short and long times,
respectively. The useful feature is just observed in CD and HN
relaxation. Consider the process $T(S_t)$ as a subordinator to
exponentially decreasing states. Let it be indexed by $\gamma$,
i.\ e. the process is obtained from a $\gamma$-stable random
process. The relaxation function for the two-state system takes
the form
\begin{equation}
\phi_{\rm CD}(t)=\int^\infty_1e^{-tz\omega_p}\,\frac{z^{-1}\,
(z-1)^{-\gamma}}{\Gamma(\gamma)\Gamma(1-\gamma)}\,dz\,.
\label{eq10}
\end{equation}
Here the subscript CD is not by chance. It shows a direct
connection of the relaxation function with the Cole-Davidson law \cite{J1}.
In fact, the one-sided Fourier transform (\ref{eq8}) gives
$$\chi_{\rm CD}(\omega)= \frac{1}{[1+i\omega/\omega_p]^\gamma}\,,
\quad 0<\gamma\leq 1\,.$$
It should be mentioned that the theory of subordination suggests
also one more scenario leading to the CD relaxation. It is based
on the inverse tempered $\alpha$-stable process (see
\cite{StanWer09} in more details).

\section{Relaxation from compound subordinators} The CC and CD
relaxations are only special cases of the more general HN law. To
get that law, the operational time $S_t=S_\alpha(t)$ of the CC
diffusion mechanism has to be modified \cite{wjmwt2010} by means
of coupling between jumps and interjump times in the underlying
continuous time random walk scheme. In other words, the temporal
decay of a given mode, representing excitation undergoing
diffusion in the relaxing system, will be characterized by short-
and long-time power laws with different fractional exponents (as
in the HN case) only if the anomalous diffusion scenario is based
on a compound operational time. To construct such an operational
time, denote conveniently the processes $T^-(S_t)$ and $T(S_t)$ as
$X^U_\gamma$ and $X^O_\gamma$, respectively. They corresponds to
the under- and overshooting subordination scenarios
\cite{wjmwt2010}. Next, we can write $Z^U_{\alpha,\gamma}(t)\leq
S_\alpha(t) \leq Z^O_{\alpha,\gamma}(t)$ for $t\geq 0$, where
$Z^U_{\alpha,\gamma}(t)=X^U_\gamma[S_\alpha(t)]$,
$Z^O_{\alpha,\gamma}(t)=X^O_\gamma[S_\alpha(t)]$. The overshooting
subordinator leads (stretching the operational time $S_\alpha(t)$)
to the HN relaxation in the form
\begin{equation}
\phi_{\rm HN}(t)=\int^\infty_1 E_\alpha\Big[-(\omega_pt)^\alpha
z\Big]\,\frac{z^{-1}\,(z-1)^{-\gamma}}{\Gamma(\gamma)\Gamma(1-\gamma)}\,dz\,.
\label{eq11}
\end{equation}
By direct calculations of the Fourier transformation (\ref{eq8})
the susceptibility reads
$$\chi_{\rm HN}(\omega)= \frac{1}{[1+
(i\omega/\omega_p)^\alpha]^\gamma}\,,\quad 0<\alpha,\gamma\leq 1\,.$$
The above approach demonstrates clearly a success in the
probabilistic treatment of the observed relaxation laws.
Therefore, we continue our analysis as applied to the
undershooting (compressing $S_\alpha(t)$) subordinator
$X^U_\gamma[S_\alpha(t)]$.

\begin{figure}
\onefigure[width=8.6 cm]{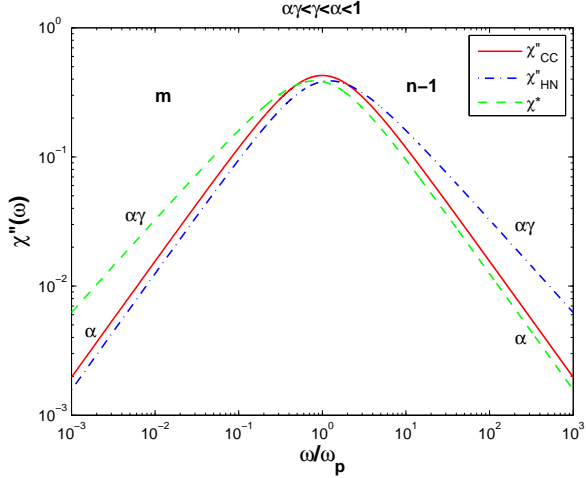} \caption{(Color online)
Imaginary term of the dielectric susceptibility for the
Havriliak-Negami, Cole-Cole relaxation and the function
$\chi^*(\omega)$ ($\alpha=0.9$, $\gamma=0.8$).}\label{fig:gml}
\end{figure}

In this case we obtain a new relaxation function
\begin{equation}
\phi(t)=\int^1_0 E_\alpha\Big[-(\omega_pt)^\alpha
z\Big]\,\frac{z^{\gamma-1}\,(1-z)^{-\gamma}}{\Gamma(\gamma)\Gamma(1-\gamma)}\,dz\,,
\label{eq12}
\end{equation}
which can be identified as a special case of the generalized Mittag-Leffler function
\cite{MatSaxHaub09}
$$E_{\alpha,\beta}^\gamma(x)=\sum_{k=0}^\infty
\frac{(\gamma,k)\,x^k}{\Gamma(k\alpha+\beta)k!}\,,\quad
\alpha,\beta>0\,,$$
where $(\gamma,k)=\gamma(\gamma+1)(\gamma+2)\dots(\gamma+k-1)$ is
the Appell's symbol with $(\gamma,0)=1$, $\gamma\neq 0$. To avoid
any confusion, it should be mentioned that the two-parameter
Mittag-Leffler function $E_{\alpha,\beta}(z)=\sum
z^k/\Gamma(k\alpha+\beta)$, more common in literature \cite{Erd},
is a special case of the function $E_{\alpha,\beta}^\gamma(z)$
with $\gamma=1$. From the series expansion of the ordinary
Mittag-Lefller function $E_\alpha(z)$ it is easy to check by
direct calculations of Eq.(\ref{eq12}) that $\phi(t)=
E_{\alpha,1}^\gamma(-(\omega_pt)^\alpha)$. This type of the
relaxation function has been derived in the continuous time random
walk framework by Jurlewicz and Weron in \cite{wjmwt2010}. Using
relation with the Mittag-Leffler function we may write now the
kinetic equation for (\ref{eq12}) in the pseudodifferential
equation form
$$\left(\frac{\partial^\alpha}{\partial t^\alpha}+
\omega_p^\alpha\right)^\gamma\phi(t)=\frac{t^{-\alpha\gamma}}{\Gamma(1-\alpha\gamma)}\,,$$
where $\partial^\alpha/\partial t^\alpha$ is the Riemann-Liouville
fractional derivative \cite{Goren}, and $\phi(0)=1$ the
initial condition. Taking the Fourier transform (\ref{eq8}), we
get the susceptibility corresponding to (\ref{eq12}) in the form
useful for fitting the dielectric spectroscopy data
\begin{equation}
\chi^*(\omega)=
1-\frac{1}{[1+(i\omega/\omega_p)^{-\alpha}]^\gamma}\,,\quad
0<\alpha,\gamma\leq 1\,. \label{eq13}
\end{equation}
This result points to the following relationship with the HN
relaxation function (see Fig.~\ref{fig:gml})
$$\chi^*(\omega)=1-(i\omega/\omega_p)^{\alpha\gamma}\chi_{\rm HN}(\omega)\,.$$
For $\gamma\to 1$ the probability density of $X^U_\gamma$ tends to
the Dirac $\delta$-function. It is easy to check $\lim_{\gamma\to
1}\phi(t)=\phi_{\rm CC}(t)$, and the kinetic equation for
$\phi(t)$ takes the Cole-Cole form mentioned above.

\begin{figure}
\onefigure[width=8.6 cm]{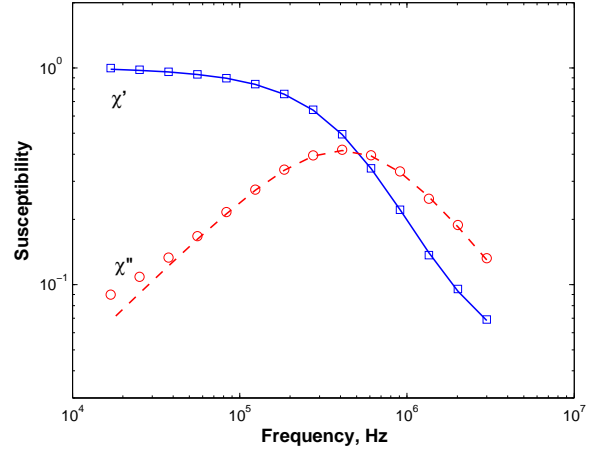} \caption{(Color online) Log-log
scale fitting example of the dielectric spectroscopy data obtained
for gallium doped Cd$_{0.99}$Mn$_{0.01}$Te mixed crystal (see more
details in the text). The straight and dotted lines represent fits
to susceptibility data points for $T=77$ K. The parameter values
$\alpha\approx0.97$, $\gamma\approx 0.78$, $\chi(0)\approx
1.0165$, $\omega_p/2\pi\approx 507$ kHz are estimated by the
method of least squares. In the frequency range 10$^4$-10$^5$ Hz
one observes superposition \cite{J2} of direct current and loss
processes indicated by a deviation of $\chi''(\omega)$ data points
from the theoretical curve.}\label{fig:fit}
\end{figure}

\section{Interpretation of experimental data} It is interesting to
compare $\chi_{\rm HN}(\omega)$ and $\chi^*(\omega)$
in the framework of experimental data results. The fractional
two-power relaxation dependencies
\begin{eqnarray}
\chi(\omega)\sim\left(i\omega/\omega_p\right)^{n-1}\quad
&{\rm for}&\quad\omega\gg\omega_p\,,\nonumber\\
\Delta\chi(\omega)=\chi(0)-\chi(\omega)\sim
\left(i\omega/\omega_p\right)^m\quad &{\rm
for}&\quad\omega\ll\omega_p\nonumber
\end{eqnarray}
are widely observed for many materials in experimental relaxation
studies \cite{J1,J2}. The exponents $n$ and $m$ fall in the range
$(0,1)$, and $\omega_p$ denotes the loss peak frequency. The HN
relaxation is characterized by the exponents $m=\alpha$ and
$m>1-n=\alpha\gamma$, and it fits the so-called typical relaxation
processes. The function $\chi^*(\omega)$ demonstrates up-down with
$m=\alpha\gamma$ and $m<1-n=\alpha$, and it fits the less typical
two-power-law relaxation pattern which, as shown by the
experimental evidence, cannot be neglected (see e.\ g.
\cite{J1,J2,hav94} and references therein). Such a less typical
behavior has been also observed by us in gallium (Ga)-doped
Cd$_{0.99}$Mn$_{0.01}$Te mixed crystals \cite{just} (see Fig.
\ref{fig:fit}, where sample frequency-domain data measured for
Cd$_{0.99}$Mn$_{0.01}$Te:Ga at 77K is fitted with the function
(\ref{eq13})). This material belongs to semiconductor of group
II-VI possessing deep metastable recombination centers. Formation
of such centers in Cd$_{\rm 1-x}$Mn$_{\rm x}$Te:Ga results from
the bistability of Ga dopant which makes this mixed crystal as an
attractive material for holography and high-density data storage
(optical memories).

In the language of
subordinators this means that the process $X^U_\gamma(t)$ makes a
rescaling for small times, and the process $X^O_\gamma(t)$ turns
on a similar rescaling for long times.
As for $0<\alpha,\gamma<1$, in the case of HN relaxation the
declination of the imaginary susceptibility $\chi''(\omega)$ for
low frequencies will be greater than for high frequencies, whereas
the less typical relaxation shows an opposite relation.

The original HN relaxation \cite{J1,J2} with exponents $0<\alpha,\gamma\leq 1$
satisfies $m\geq1-n$. Its modified version \cite{hav94}, proposed
to fit relaxation data with power-law exponents satisfying $m<1-n$,
assumes  $0<\alpha,\alpha\gamma\leq 1$. Unfortunately, the HN function
with $\gamma > 1$ cannot be derived within the framework of
diffusive relaxation mechanisms. Only for $\gamma\leq 1$ the origins
of the HN function can be found within the fractional Fokker-Planck
\cite{kalm} and continuous time random walk \cite{wjmwt2010} approaches.
The approach considered above includes all the data in one, mathematically
unified approach.

\section{Conclusions} We have discovered a novel law of relaxation
accompanied by the well-known Havriliak-Negami function. Earlier
the development of the theory of nonexponential relaxation went
behind the fitting of experimental data. Now the subordination
approach allows one to explain not only the well-known experimental
laws of relaxation, but it makes a prediction of other adequate
models useful for experimentalists.

\acknowledgments AS is grateful to the Institute of Physics and
the Hugo Steinhaus Center for Stochastic Methods for pleasant
hospitality during his visit in Wroc{\l}aw University of
Technology.

\end{document}